\documentclass[12pt]{article}
\newcommand{\Z}{Z\!\!\!Z}
\newcommand{\beq}{\begin{equation}}
\newcommand{\eeq}{\end{equation}}
\newcommand{\beqn}{\begin{eqnarray}}
\newcommand{\eeqn}{\end{eqnarray}}
\newcommand{\bea}[1]{\beq\begin{array}{#1}}
\newcommand{\eea}{\end{array}\eeq}

\newcommand{\eq}[1]{(\ref{#1})}

\newcommand{\dual}[1]{{}^{*}{#1}}

\newcommand{\Tr}[1]{\;{1\over #1}\mathop{\rm Tr}}
\newcommand{\tr}{\mathop{\rm Tr}}

\newcommand{\Pexp}{\mbox{P}\!\exp}

\newcommand{\diff}{\partial}
\newcommand{\cC}{{\cal C}}

\newcommand{\cD}{{\cal D}}
\newcommand{\cL}{{\cal L}}

\newcommand{\cS}{{\cal S}}
\newcommand{\cV}{{\cal V}}

\newcommand{\NP}[3]{{\it Nucl. Phys. }{\bf #1} (#2) #3}
\newcommand{\NPPS}[3]{{\it Nucl. Phys. Proc. Suppl. }{\bf #1} (#2) #3}
\newcommand{\PL}[3]{{\it Phys. Lett. }{\bf #1} (#2) #3}
\newcommand{\PRL}[3]{{\it Phys. Rev. Lett. }{\bf #1} (#2) #3}
\newcommand{\PRep}[3]{{\it Phys. Rep. }{\bf #1} (#2) #3}
\newcommand{\PR}[3]{{\it Phys. Rev. }{\bf #1} (#2) #3}
\newcommand{\JL}[3]{{\it JETP Lett. }{\bf #1} (#2) #3}

\newcommand{\MPL}[3]{{\it Mod. Phys. Lett. }{\bf #1} (#2) #3}
\newcommand{\IJMP}[3]{{\it Int. J. Mod. Phys. }{\bf #1} (#2) #3}

\newcommand{\PTPS}[3]{{\it Prog. Theor. Phys. Suppl. }{\bf #1} (#2) #3}

\date{}
\begin{document}
\title{ Dirac Strings and Monopoles \\
in the Continuum Limit of SU(2) Lattice Gauge Theory.
\vskip-50mm
\rightline{\small ITEP-TH-7/00}
\rightline{\small MPI-PhT/2000-12}
\vskip 50mm
}
\author{M.N.~Chernodub$^{\rm a}$,
F.V.~Gubarev$^{\rm a,b}$, M.I.~Polikarpov$^{\rm a}$,\\
V.I.~Zakharov$^{\rm b}$ \\
\\
$^{\rm a}$ {\small\it Institute of Theoretical and  Experimental Physics,}\\
{\small\it B.Cheremushkinskaya 25, Moscow, 117259, Russia}\\
$^{\rm b}$ {\small\it Max-Planck Institut f\"ur Physik,}\\
{\small\it F\"ohringer Ring 6, 80805 M\"unchen, Germany} \\
\vspace{2\baselineskip}
}

\maketitle
\thispagestyle{empty}
\setcounter{page}{0}
\begin{abstract}
\noindent
Magnetic monopoles are known to emerge as leading non-perturbative fluctuations
in the lattice version of non-Abelian gauge theories in some gauges. In terms of
the Dirac quantization condition, these monopoles have magnetic charge $|Q_M|=2$.
Also, magnetic monopoles with $|Q_M|=1$ can be introduced on the lattice via the
't~Hooft loop operator. We consider the $|Q_M|=1,2$ monopoles in the continuum
limit of the lattice gauge theories. To substitute for the Dirac strings which cost no
action on the lattice, we allow for singular gauge potentials which are absent in the
standard continuum version. Once the Dirac strings are allowed,
it turns possible to find a solution with zero action for a monopole--antimonopole
pair. This implies equivalence of the standard and modified continuum versions in
perturbation theory. To imitate the nonperturbative vacuum, we introduce then
a nonsingular background. The modified continuum version of the gluodynamics
allows in this case for monopoles with finite non-vanishing action.
Using similar techniques, we construct the 't~Hooft loop operator in the continuum and
predict its behavior at small and large distances both at zero and high temperatures.
\end{abstract}

\newpage
\section*{Introduction}
\noindent
While perturbative Yang-Mills theories appear to be understood beyond any doubt,
non-perturbative physics is much more challenging at the moment. Moreover, the
main source of knowledge in the non-perturbative domain is the lattice gauge
theories. In particular, there exist rich data  supporting  the idea of the
quark confinement through the magnetic monopole condensation (for review,
see, e.g.~\cite{reviews}).

Any analytical treatment of magnetic monopoles in the continuum limit represents
apparent difficulties because of singularities in the gauge potential $A_{\mu}$.
Indeed, such singularities are displayed already by the original Dirac monopole:
\beq
\label{dirac}
A_\mu dx_\mu = {1\over 2}\;(1 + \cos\theta)\;d\varphi,
\eeq
or in the component form in the spherical coordinates: 
\beq
A_{\theta}=A_r=0, \qquad
A_{\varphi}= {1\over 2}\;{(1+\cos\theta)\over r \sin\theta}\;.
\eeq
The singularity along the line $\theta=0$ represents the Dirac string, while the
singularity at $r\to 0$ corresponds to a singular  magnetic filed,
${\bf H} \sim {\bf r}/r^3$. In non-Abelian theories with Higgs mechanism the
singularities are resolved and there exists the famous 't~Hooft-Polyakov
solution~\cite{TP-solution} with finite energy. In a particular gauge, the
corresponding potential is given by
\beq
\label{hp}
A^a_i~=~f(r){\varepsilon_{aik}r_k\over  r^2}
\eeq
where $a$ is the color index, $a=1,2,3$ and $f(r)\to 0$ as $r\to 0$ while $f(r)\to 1$
as $r\to \infty$.

In pure gauge theories, there are no monopole solutions with finite
energy. To reconcile this with observation of monopoles on the
lattice, one considers dual gauge theories which serve as infrared
limit of QCD \cite{baker}. In its simplest version, the theory is
build on an octet of dual gluons and three octets of scalar (Higgs)
fields. In this paper, we would stick to consideration of monopoles
within the fundamental QCD. The reason is that the monopoles on the
lattice are defined beginning from elementary cubes, i.e. at smallest
distances available.  Our guiding principle is to reexamine the
continuum limit by confronting the treatment of the
monopole-associated singularities on the lattice and in the continuum.

In the lattice formulation, the singularities due to the Dirac string
and at $r\to 0$ are treated differently. As was emphasized first by
Polyakov~\cite{Polyakov-compact-U1}, the Dirac strings are allowed,
i.e. cost no action in the lattice compact U(1) theory. As for the $r\to
0$ singularity, it introduces in this case a physical divergence in
the action. The suppression due to this divergence is overcome,
however, by the entropy factor when the coupling constant $g$,
included into the definition of $A_{\mu}$ above, is of order of unity.

In the non-Abelian gauge models the relation of monopoles to the
action is much more obscure, as far as analytical results are
concerned. Moreover, one of the most important steps in introducing
monopoles is a pure topological definition which makes no reference to
the associated non-Abelian action~\cite{tHooft-abelian-gauges}. In
this formulation, monopoles are related to topology of gauge
fixing. Namely, if the gauge is fixed (up to U(1) rotations) by
directing a color vector $h^a$ in, say, the third direction, then the
fixation fails at the points where all the components $h^a$ vanish.
Moreover, one can prove that such points belong to monopole
trajectories. The function of $h^a$ can be played by any vector, for
example, by a particular Lorenz component of the gluonic
field-strength tensor, say, $F_{12}^a$. Vanishing of $h^a$ has no
direct effect on the action.  Other Abelian projections revealing
monopoles are also known, the most famous one seems to be the Maximal
Abelian gauge (for review and references see \cite{reviews}).

Monopoles which condense in the confining phase have magnetic charges
$|Q_M|=2$, the same as the 't~Hooft-Polyakov monopole (\ref{hp}). In
the corresponding U(1) projection the associated Dirac string does not
introduce infinities because of the compactness of the U(1) subgroups
of SU(2), see the discussion above. On the other hand the Dirac
strings associated with $|Q_M|=1$ monopoles are not allowed in the QCD
vacuum since in the continuum limit they have infinite
energy. However, one can introduce $|Q_M|=1$ monopoles as external
objects via the 't~Hooft loop operator~\cite{tHooft-tHooft-loop}.

In this paper, we consider magnetic monopoles in the continuum provided that the continuum
is understood as the limiting case of lattice theories. First, we generalize the treatment of
Dirac strings within the lattice compact QED to the non-Abelian case. As expected, the lattice
formulation of the non-Abelian theories corresponds to non-observability of the Dirac strings,
defined in a particular way. To substitute for their effect in the continuum, one allows for
certain singular potentials. Thus, we argue that the standard continuum formulation is to be
modified in a certain way to allow for the Dirac strings.

It is amusing that once the Dirac strings are admitted into the
continuum limit the $|Q_M|=2$ monopoles cost no action either. Namely, we
construct an explicit solution with zero action for a Dirac strings
with open ends. In this respect the non-Abelian theories differ
radically from their Abelian counterpart where the end points of the
Dirac strings represent monopoles (\ref{dirac}) with divergent
action. It might worth emphasizing that the Abelian part of the fields
in the no-action solution does correspond to the standard Abelian
monopoles and it is the commutator term in the field-strength tensor
which allows to nullify the non-Abelian action. This is in the
correspondence with the instability of a single $|Q_M|=2$ monopole
with nonzero action in the non-Abelian pure gauge theory which is
known since long \cite{Brandt-Neri}.

The explicit monopole-pair solution with no action mentioned above is
obtained in empty, or perturbative vacuum. We check that quantum
fluctuations around this zero-action solution do not distinguish it
from the perturbative vacuum either. Therefore the modified continuum
version corresponding to the limit of the lattice theories brings no
change in the perturbative domain as compared with the standard
Lagrangian theory.

We then imitate non-perturbative vacuum of QCD by introducing
nonsingular background fields, $F_{\mu\nu}^{soft}\sim
\Lambda^2_{QCD}$. Then there still exist Dirac strings with zero
action whose color orientation is aligned with that of the background
field. On the other hand, introduction of monopoles a la 't~Hooft (see
above) is related to some singular gauge transformations with their
own color orientation. As a result, monopoles in the physical vacuum
are associated, generally speaking, with an action of order
$L\cdot\Lambda_{QCD}$ where $L$ is the length of the monopole world
trajectory.

Finally, the same techniques as used to construct invisible Dirac
strings in the continuum limit produce a continuum analog for the
't~Hooft loop operator. It shares the basic properties of the 't~Hooft
loop operator and allows to formulate new predictions for the
intermonopole potential.  At short distances the 't~Hooft loop
describes the Coulomb-like interaction of the monopoles with
$|Q_M|=1$, Ref. \cite{Samuel-83}. We fix the coefficient at front of
this Coulombic term.  At larger distances the $|Q_M|=1$ monopoles,
introduced via the 't~Hooft loop interact with the $|Q_M|=2$ monopoles
of the medium. We describe this interaction within the effective
Abelian Higgs model (for review and references see \cite{reviews}),
which uniquely fixes the Yukawa-like behavior of the intermonopole
potential. We include also consideration of the 't~Hooft loop at high
temperatures where the Debye screening becomes essential.
 
The outline of the paper is as follows. In Section~\ref{symmetriesSc}
we show that symmetries of the lattice and standard continuum actions
of gluodynamics are different. We propose the modified continuum
action which allows for the Dirac strings.  In
Sections~\ref{AbelianGaugeSc}, \ref{Strings-in-background} monopole
configurations within the new approach are considered.  In
Section~\ref{tHooftSc} we introduce the 't~Hooft loop operator in the
continuum. In Section~{\ref{Monopole-interaction}} the predictions for
the 't~Hooft loop are formulated. Our conclusions are summarized in
the last section.

\section{Dirac Strings in SU(2) Gauge Theory.}
\label{symmetriesSc}
\noindent
As is mentioned in the Introduction, the lattice formulation of the
compact photodynamics gives a version of the U(1) gauge theory
\cite{Polyakov-compact-U1} with unobservable Dirac strings. In this
section we develop a generalization of this construction to the case
of SU(2) gauge model.

The general one--plaquette action of SU(2) lattice gauge theory (LGT) can be represented as:
\beq
\label{LGT-general}
S_{lat} (U) = \frac{4}{g^2} \sum\limits_{p} S_P(1-\Tr{2} U[\diff p])\,,
\eeq
where $g$ is the bare coupling, $\diff p$ is the boundary of an
elementary plaquette $p$, the sum is taken over all $p$, $U[\diff p]$
is the ordered product of link variables $U_l$ along $\diff p$. To
have the correct naive continuum limit the function $S_P$ should obey
the condition $\lim\nolimits_{x\to 0} S_P(x) = x + \cdots\;$. In
particular, if $S_P(x) = x$ then (\ref{LGT-general}) is the standard
Wilson action.  The exponent of the lattice field strength tensor
$F_p$ defines $U[\diff p]$:
\beq
U[\diff p] ~=~ e^{  i \hat F_p } ~=~
\cos[\frac{1}{2}|F_p|]\; + i \tau^a n^a_p \; \sin[\frac{1}{2}|F_p|]\,,
\eeq
where $\hat F = F^a \cdot \tau^a/2$, $|F| = \sqrt{F^a F^a}$ and we
define $n^a_p = F^a_p/|F_p|$ for $|F_p| \neq 0$, $n^a$ is an arbitrary unit
vector for $|F_p| = 0$. Sometimes we also use the vector-like notations
$\vec{F}$ instead of $F^a$. The lattice action (\ref{LGT-general})
depends only on $\cos[\frac{1}{2}|F_p|]$.  Therefore the action of the
SU(2) LGT possesses not only the usual gauge symmetry, but allows also
for the gauge transformations which shift the field strength by $4\pi
k$, $|F_p| \to |F_p| + 4\pi k$, $k \in Z$:
\bea{ccl}
\label{lattice-discrete-symmetry-1}
\begin{array}{lll}
e^{ i \hat F_p }  &  =  &  \exp\{ \; i |F_p| \; {\hat n}_p \; \} \; = \\
& = &
\exp\{ \; i (|F_p| + 4\pi) \; {\hat n}_p\; \} =
\exp\{ \; i (F^a_p + 4\pi n^a_p ) \; \tau^a/2 \; \}\,,
\end{array}
\eea
Thus the symmetry inherent to the lattice formulation can be represented as:
\beq
\label{lattice-discrete-symmetry-2}
F^a_p ~ \to ~ F^a_p + 4\pi n^a_p\,,
\qquad
\vec{F}_p \times \vec{n}_p = 0\,,
\qquad
\vec{n}_p^2 = 1\,.
\eeq
The symmetry (\ref{lattice-discrete-symmetry-2}) is absent in the conventional continuum action,
$\int (F^a_{\mu\nu})^2\, d^4 x$ and therefore the continuum limit of SU(2) LGT is different from
the commonly accepted SU(2) gluodynamics at least in this respect. Below we explore the
consequences of Eq.~(\ref{lattice-discrete-symmetry-2}) for the continuum theory.

In the continuum limit $n^a_p$ becomes a singular two-dimensional structure
$\dual \Sigma^a_{\mu\nu} = \frac{1}{2} \varepsilon_{\mu\nu\lambda\rho} \Sigma^a_{\lambda\rho}$
which is a generalization of the Dirac strings in the compact electrodynamics and which
transforms in the adjoint representation of the gauge group. Consider first a special class of
the gauge potentials which may be gauge transformed to pure Abelian fields,
$A^a_\mu = \delta^{a,3} A_\mu$. For such fields the action of the SU(2) gluodynamics
coincides with the action of the compact U(1) gauge model, up to the ghost terms.
Therefore in this gauge $\Sigma^a_{\mu\nu} = \delta^{a,3} \Sigma_{\mu\nu}$, where
$\Sigma_{\mu\nu}$ is nothing else but the Dirac string:
\beq
\label{Sigma}
\Sigma_{\mu\nu} = \int d^2\sigma \sqrt{g} \; t_{\mu\nu}(\sigma) \;
\delta^{(4)}(x-\tilde{x}(\sigma))\,,
\eeq
with the world-sheet coordinates $\tilde{x}(\sigma)$ parameterized by $\sigma_\alpha$,
$\alpha=1,2$:
\beq \label{tg}
t_{\mu\nu}(\sigma) =  {1\over \sqrt{g}} \; \varepsilon^{\alpha\beta} \;
\diff_\alpha \tilde{x}_\mu \; \diff_\beta \tilde{x}_\nu\,,
\qquad
t^2_{\mu\nu} = 2\,,
\qquad
g(\sigma) = \mathrm{Det}[\; \diff_\alpha \tilde{x}_\mu \;
\diff_\beta \tilde{x}_\mu \;]\,.
\eeq
Thus for general gauge potentials
\beq
\label{Sigma-colored-1}
\Sigma^a_{\mu\nu} = \int d^2\sigma \sqrt{g} \; t^a_{\mu\nu}(\sigma) \;
\delta^{(4)}(x-\tilde{x}(\sigma))\,.
\eeq
The second equality in (\ref{lattice-discrete-symmetry-2}) requires that
\beq
\label{constraint-1}
\vec{t}_{\mu\nu}(\sigma) \times \dual{\vec{F}}_{\mu\nu}(\tilde{x}) = 0\,,
\eeq
where the continuum field strength tensor ${\hat F}_{\mu\nu} = \diff_\mu {\hat A}_\nu - 
\diff_\nu {\hat A}_\mu - i [{\hat A}_\mu, {\hat A}_\nu]$. Eq.~(\ref{constraint-1})
determines the color structure of $t^a_{\mu\nu}$:
\beq
\label{tn}
t^a_{\mu\nu}(\sigma) ~=~ t_{\mu\nu}(\sigma) ~ n^a(\sigma) \, ,
\qquad
n^a(\sigma) ~=~ (t \cdot \dual{F}^a ) \left[{(t \cdot \dual{F}^b )^2}\right]^{-1/2}
\,,
\eeq
where $(t \cdot F^a) \equiv t_{\mu\nu}(\sigma) \;
F^a_{\mu\nu}(\tilde{x})$ and $n^a$ is normalized as $\vec{n}^2=1$. On
the set of points where $(t \cdot \dual{F}^a) = 0$ the direction of
$n^a(\sigma)$ is arbitrary. Therefore, in the general case
$\Sigma^a_{\mu\nu}$ is given by
\beqn
\Sigma^a_{\mu\nu} ~=~ \int d^2\sigma \; \sqrt{g}\; t_{\mu\nu}(\sigma) \; n^a(\sigma) \;
\delta^{(4)}(x-\tilde{x}(\sigma))\,,
\nonumber \\
\label{Sigma-colored} \\
\vec{n}^{\;2}(\sigma)~=~1\,,
\qquad
\vec{n}(\sigma) \times (t_{\mu\nu}(\sigma)\;\dual{ \vec{F}}_{\mu\nu}(\tilde{x})) ~=~ 0\,.
\nonumber
\eeqn
and the continuum analog of the lattice symmetry
Eq.~(\ref{lattice-discrete-symmetry-1},\ref{lattice-discrete-symmetry-2}) is:
\beq
\label{discrete-symmetry}
F^a_{\mu\nu} ~ \to ~ F^a_{\mu\nu} + 4\pi \; \dual{\Sigma}^a_{\mu\nu}\,,
\eeq

Note that we do not claim that the only string-like singularities which may exist in the
continuum limit of SU(2) LGT are of the type (\ref{Sigma-colored},\ref{discrete-symmetry}).
Indeed, there are known examples of various Abelian gauges (see \cite{reviews} and
references therein) in which Abelian monopoles and Dirac strings naturally arise. String singularities
in these gauges are of the type (\ref{Sigma-colored}), but their color orientation is different.
Therefore the strings (\ref{Sigma-colored}) are not the most general. Nevertheless, we claim that only
the strings (\ref{Sigma-colored}) produce no additional action. In other words the action of the
SU(2) LGT in the continuum limit calculated with $F^a_{\mu\nu}$ and
$F^a_{\mu\nu}+4\pi \dual{\Sigma}^a_{\mu\nu}$  is the same only if $\Sigma^a_{\mu\nu}$ is given by
(\ref{Sigma-colored}). We shall come back to discuss this issue in Section~\ref{Strings-in-background}.

The action of SU(2) gluodynamics which possesses the additional symmetry (\ref{discrete-symmetry})
can be formally represented as:
\beq
\label{part-fun-general}
Z= \int \cD A\;\exp\Bigl\{ - S(F) \Bigr\}\,,
\eeq
\beq
S(F) = - \log \; \int \cD\Sigma \exp\Bigl\{ - {1\over 4 g^2} \int d^4 x \;
\Bigl[ F^a_{\mu\nu} + 4\pi\; \dual{\Sigma}^a_{\mu\nu} \Bigr]^2
\Bigr\}\,,
\label{new-action}
\eeq
where the integration is over all possible surfaces (\ref{Sigma-colored}). The expressions
(\ref{part-fun-general},\ref{new-action}) are only formal since, as we show in
Section~\ref{AbelianGaugeSc} it is impossible to separate rigorously the measure $\cD\Sigma$
from the gauge degrees of freedom in $\cD A$.  Nevertheless, the
Eq.~(\ref{part-fun-general},\ref{new-action}) is a good starting point for the analysis 
of the next section. Note that the action (\ref{new-action}) is invariant under smooth SU(2) gauge
transformations since vector $n^a$ transforms in the same way as $F^a_{\mu\nu}$ does. By
construction, this action is also invariant under transformations
(\ref{Sigma-colored},\ref{discrete-symmetry}) which correspond to the lattice symmetry relations
(\ref{lattice-discrete-symmetry-1},\ref{lattice-discrete-symmetry-2}).

Note also that for self-intersecting surface  $\Sigma_{\mu\nu}$, Eq.~(\ref{Sigma-colored}), the
world-sheet vector field $n^a(\sigma)$ is generally multi-valued as function of $\tilde{x}$. Furthermore,
for the non-orientable surfaces the field $n^a(\sigma)$ cannot be defined smoothly everywhere on
$\Sigma$. To avoid these complications we consider only the orientable surfaces without
self-intersections. This reservation is specific for Dirac strings in the non-Abelian case.

\section{Monopoles.}
\label{AbelianGaugeSc}
\noindent
We proceed now to consider Dirac strings with open ends. The end points can
be associated, as usual, with monopoles.  If one follows only the Abelian-like part of the field
strength tensors, these are standard Dirac monopoles. However, the full non-Abelian action is no
longer bounded in terms of the Abelian magnetic field and we will present a zero-action
solution for open Dirac strings. Therefore contrary to the Abelian models the open Dirac strings
in SU(2) gluodynamics are the gauge copies of the vacuum $A=0$. Moreover, we show that this result is
also valid at the one loop level and hence the Dirac strings (\ref{Sigma-colored}) do not change the
perturbation theory.

\subsection{String Independence.}
\label{String-independence}
\noindent
Consider the partition function~(\ref{part-fun-general},\ref{new-action}) in case of a single
surface $\Sigma^a_{\mu\nu}$:
\beq
\label{Sigma-path-integral}
Z[\Sigma] = \int \cD A \exp\Bigl\{ - {1\over 4 g^2} \int d^4 x \;
\Bigl[ F^a_{\mu\nu} + 4\pi q \; \dual{\Sigma}^a_{\mu\nu}
\Bigr]^2 \Bigr\}\,,
\eeq
where the constant $q$ is equal to unity in (\ref{part-fun-general},\ref{new-action}). Varying the
gauge fields $A$ we get the classical equations of motion:
\beq
\label{Sigma-eq-motion}
D_\nu \left( \hat F_{\mu\nu}(A) + 4\pi q \; \dual{\hat \Sigma}_{\mu\nu}\right) = 0\,,
\eeq
which should be supplemented by Bianchi identities:
\beq
D_\nu \dual{\hat F}_{\mu\nu} = 0\,.
\eeq
Note that Eq.~(\ref{Sigma-eq-motion}) is consistent with the covariant conservation of
electric currents:
\beqn
D_\mu D_\nu \, \hat F_{\mu\nu} = - 4\pi q \, D_\mu D_\nu \, \dual{\hat \Sigma}_{\mu\nu}
\sim \int d^2\sigma_{\mu\nu} \dual{\left[ D_\mu , D_\nu \right] \hat n(\sigma)}
\, \delta^{(4)} (x - {\tilde x}(\sigma)) = 0\,.\nonumber
\eeqn
where the last equality is due to (\ref{Sigma-colored}).

To appreciate the meaning of eq.\eq{Sigma-eq-motion}
let us confine ourselves for the moment  to the fields $A^a_\mu = Q^a \cdot A_\mu$
with a constant color direction $Q^a$. Then $n^a \sim Q^a$ and Eq.~(\ref{Sigma-eq-motion}) becomes:
\beq
\label{abelian-equations}
\diff_\nu \left( \diff_{[\mu} A_{\nu]} \right) =
- 4\pi q \; \diff_\nu \dual{\Sigma}_{\mu\nu}\,.
\eeq
The solution of this equation in the Landau gauge,
\beq
A^a_\mu = - Q^a \cdot 4\pi q
\; {1\over \Delta}\; \diff_\nu \dual{\Sigma}_{\mu\nu}\,,
\eeq
corresponds to the gauge potential of an Abelian monopole current $\diff \Sigma$ embedded into
the SU(2) group. Thus $\Sigma_{\mu\nu}$ is the Dirac string worldsheet.

Let us show that the shape of $\Sigma$ is irrelevant, that is the surface  $\Sigma$ can be shifted by
a gauge transformation  provided that the boundary $\diff\Sigma$ is fixed.  Assuming that the
orientable surface (\ref{Sigma-colored}) has no self-intersections, we may write
$\Sigma^a_{\mu\nu} = n^a \Sigma_{\mu\nu}$. Consider then a closed non self-intersecting surface $\cS$ on
which the vector field $n^a(\sigma)$ is a single valued function of $\tilde{x}$.
We can define the field $n^a(x)$, $\vec{n}^{\;2}=1$, in the whole space-time in such a way that
\beq
\label{n-a-on-surface}
n^a(x) = n^a(\tilde{x})\quad \mbox{for} \quad x\in \cS \,.
\eeq
Note that the definition of the vector $n^a(x)$ is not unique, but this is irrelevant for our analysis.

Consider the following gauge transformation matrix:
\beqn
\label{closed-gauge-trans}
\Omega (\cV_\cS) & = & \exp\{\; i \; \alpha(\cV_\cS, x) \; \vec{n}(x)\vec{\tau} \;\}\,,
\\
\label{closed-gauge-trans-param}
\alpha(\cV_\cS, x) & = & 2\pi q \;\int\limits^x_\infty V_\mu dx_\mu\,, \qquad
V_\mu = \int\limits_{\cV_\cS} \left(\dual{d^3\zeta}\right)_\mu \delta^{(4)}(x-\zeta)\,,
\eeqn
where $V_\mu$ is a characteristic function of the volume $\cV_\cS$ bounded by the surface
$\cS$. The first integral in (\ref{closed-gauge-trans-param}) is taken along any path $C_x$ connecting
infinity with the point $x$. Under the  general gauge transformation $\Omega$ the field strength
tensor transforms as:
\beq
\label{gauge-trans-F-mu-nu}
\hat F_{\mu\nu}( A^\Omega) =
\Omega^+ \hat F_{\mu\nu}(A) \Omega + i \Omega^+ \left[\diff_\mu, \diff_\nu \right] \Omega =
\Omega^+ \hat F_{\mu\nu}(A) \Omega + \hat F_{\mu\nu}(i\Omega^+\diff\Omega)\,.
\eeq
Straightforward calculations show that for $\Omega$ defined by Eq.~(\ref{closed-gauge-trans})

\beqn
F^a_{\mu\nu}(i\Omega^+\diff\Omega) \, & = & \, -2 n^a \; [\diff_\mu, \diff_\nu] \alpha \\
& & 
-\left( \sin[2\alpha] \; \delta^{ac} + (1-\cos[2\alpha]) \varepsilon^{abc} n^b
\right) \, [\diff_\mu, \diff_\nu] n^c\,.
\nonumber
\eeqn
Note that the function $\alpha(x)$ takes only two values, $0$ and $2\pi q$.
Therefore for integer or the half-integer valued charge $q$ we have:
\beq
\label{D-mu-nu-Omega-1}
F^a_{\mu\nu}(i\Omega^+\diff\Omega) = - 2 \; n^a \; [\diff_\mu, \diff_\nu] \alpha =
- 4\pi q \; n^a \; \diff_{[\mu} V_{\nu]} = - 4\pi q \; n^a \; \dual{\cS}_{\mu\nu}\, .
\eeq
Thus the gauge transformation considered adds a closed surface
$\dual{\cS}_{\mu\nu}$ to the field strength tensor, $\hat{F}(A^\Omega)
= \Omega^+ \hat{F}(A) \Omega - 4\pi q \; \dual{\hat{\cS}}$, $\hat{\cS}
= \hat{n} \cS$. It is easy to see that the color structure of the
surface $\Sigma^a_{\mu\nu}$ was inessential in our analysis. Indeed,
one may perform exactly the same transformations with arbitrary
$n^a(\sigma)$, $\vec{n}^{\;2}=1$, instead of (\ref{Sigma-colored}).
Therefore the orientable non self-intersecting surface $\Sigma$ in
Eq.~(\ref{Sigma-path-integral}) with arbitrary color orientation can
be deformed by the singular gauge transformation provided that $q= 0,
\pm \frac{1}{2}, \pm 1, ...$.

We see that the situation looks similar to the Abelian case where the shape of the Dirac strings is
inessential and can be changed by a gauge transformation so that only the end points of the strings
have a physical meaning: they are identified with monopoles. However despite of
this similarity the Yang--Mills theory is different in some respects. In particular, for the examples
considered below the boundaries of the strings (\ref{Sigma-colored}) have zero action thus being a
pure gauge artifacts\footnote{
Note that in the
standard Yang--Mills theory these configurations have infinite
action and therefore are not important.
}.

\subsection{Open Strings With Zero Action.}
\noindent
Consider the following gauge transformation matrix:
\beq
\label{kyoto-omega-1}
\Omega_1 = \left(
\begin{array}{cc}
\cos{\theta\over 2} \; e^{i\varphi} & \sin{\theta\over 2} \\
 & \\
-\sin{\theta\over 2}  & \cos{\theta\over 2} \; e^{-i\varphi}
\end{array}
\right)\,,
\qquad\qquad
\Omega_1^+ \tau^3 \Omega_1 = \hat{x}^a \tau^a\,,
\eeq
defined in the time-slice $t=0$; $~\theta$, $\varphi$ are polar and azimuthal angles.  Performing the
gauge transformation on the pure vacuum configuration $A=0$ one gets:
\beq
\label{kyoto-omega-1-field-strength}
\dual{F}^a_{\mu\nu}( A^{\Omega_1}) = 4\pi \; \delta_{0,[\mu} \; \delta_{\nu],3} \;
\delta^{a,3} \cdot \Theta(z) \delta(x)\delta(y)\,.
\eeq
Therefore the singular gauge transformation (\ref{kyoto-omega-1})
produces singular $F^a_{\mu\nu}$ as well, but the singularity in
(\ref{kyoto-omega-1-field-strength}) is of allowed type
(\ref{discrete-symmetry}) with time independent surface $\Sigma$
directed along $\tau^3$ in the color space. On the other hand, in the
gauge transformed potentials one finds an Abelian monopole which is
double charged in terms of the minimal Dirac quantization condition (cf. Eq.~(\ref{dirac})):
\bea{ccl}
\label{kyoto-monopole}
A^3 & = & A^3_\mu dx_\mu = - (1 + \cos\theta ) \; d\varphi\,,\\
\rule[-3mm]{0mm}{7mm}
A^+ & = & (A^1_\mu + i A^2_\mu) dx_\mu =
- e^{i\varphi}\;(d\theta - i \sin\theta\; d\varphi)\,.
\eea
The interpretation of (\ref{kyoto-omega-1-field-strength},\ref{kyoto-monopole}) is as follows.
In the U(1) case due to the magnetic flux conservation the Dirac string terminates at an Abelian
monopole with the magnetic field $|{\bf H}|\sim 1/r^2$. In the SU(2) gauge model the Abelian string
with net flux $4\pi$ may disappear into the vacuum. Although we still have the conservation of Abelian
flux, this does not imply any bound on the action.
In fact, because of the nontrivial components $A^\pm$ the full SU(2) action is zero.  The zero action of
the configuration (\ref{kyoto-monopole}) is due to the cancellation between the Abelian-like and
commutator pieces in $F^a_{\mu\nu}$.
Note that already in Ref.~\cite{Brandt-Neri} it was shown that 
the double charged Abelian monopoles being immersed into SU(2) gauge group are unstable against
fluctuations of non-Abelian components of gauge fields.

Proceed now to generalizing (\ref{kyoto-omega-1}) to the case of finite Dirac string. Consider the
potential $A_\mu$ which in U(1) theory represents the mo\-no\-po\-le--anti\-mo\-no\-po\-le pair located
at $x,y=0$, $z = \pm R/2$:
\beq
\label{kyoto-omega-2-abelian-monopole-pair}
A_\mu dx_\mu = {1\over 2}\left( {z_+ \over r_+} - {z_- \over r_-}\right) \; d\varphi 
= A_D(z,\rho) d\varphi\,, \qquad\qquad
0 \leq A_D(z,\rho) \leq 1
\eeq
\beq
z_\pm = z \pm R/2\,, \qquad \qquad
\rho^2 = x^2 + y^2\,, \qquad \qquad
r^2_\pm = z^2_\pm + \rho^2\,,
\eeq
and the following gauge transformation matrix
\beq
\label{kyoto-omega-2}
\Omega_2 = \left(
\begin{array}{ll}
e^{i\varphi} \sqrt{A_D}  & \sqrt{1-A_D} \\
 & \\
- \sqrt{1-A_D} & e^{- i\varphi} \sqrt{A_D}
\end{array}
\right)\,.
\eeq
It is easy to check that (\ref{kyoto-omega-2}) when applied to the vacuum $A=0$ produces a string of
the type (\ref{Sigma-colored},\ref{discrete-symmetry}) which begins and terminates at the points
$\rho=0$, $z=z_\pm$, respectively:
\beq
\label{kyoto-omega-2-field-strength}
\dual{F}^a_{\mu\nu} =  4\pi \;  \delta_{0,[\mu} \; \delta_{\nu],3} \; \delta^{a,3} \cdot \Theta(R/2 - |z|) \;
\delta(x)\; \delta(y)\,.
\eeq
The corresponding gauge potential $A = i \Omega^+_2 \diff \Omega_2$ contains Abelian monopole and
antimonopole located at the ends of the string, $A^3_\mu dx_\mu= -2 A_D(z,\rho) d\phi$.

The described above monopole configurations might be related to the mo\-no\-po\-les common to the
the lattice Abelian projections \cite{reviews}. In a way, it is a consequence of the asymptotic freedom
alone. Indeed, by the monopole one understands field configurations which in their Abelian part look like
the standard Dirac monopole (\ref{dirac}). The action associated with the Abelian monopole is linearly divergent
in the ultraviolet,
$$
S_{Abelian} \sim (a g^2(a))^{-1}\,,
$$
where $a$ is an ultraviolet cut off, say, the lattice spacing. At first sight, on the background of this linear
divergence the logarithmic behavior of the coupling is not important at all.  However, it was shown in
Ref.~\cite{Polyakov-compact-U1} that the $a^{-1}$ factor in the action can be overcome by the entropy
since it is proportional to an exponential of the length of monopole trajectories measured  in the same units
of $a$. As a result, the value of the coupling is becoming crucial and the Abelian-like  monopoles can be abundant in 
the vacuum only if the coupling is of order unit, $g^2\sim 1$. Which is inconsistent with the asymptotic freedom of
the gluodynamics. The only way out is to have the non-Abelian field strength vanishing at short distances,
$F^a_{\mu\nu}\to 0$ at $r\to 0$. In other words, the cancellation of the Abelian-like and commutator terms in the
field strength tensor should be {\it exact} at short distances. The latter condition is satisfied
by (\ref{kyoto-monopole}) which appears to be the unique monopole solution at short distances

The monopole structure at short distances can be studied directly
on the lattice. At the distances available so far, the monopoles in SU(2) LGT are associated with a
sizeable excess in the action, although the excess is substantially smaller than it would be in the
pure Abelian case \cite{monopoles-last,Suganuma-98}. Further measurements at smaller distances would be very
interesting.

\subsection{Quantum Corrections.}
\noindent
The examples presented above show that an arbitrary (non self-intersecting) string 
(\ref{Sigma-colored}) may be considered as a result of combined gauge transformations of the type
(\ref{closed-gauge-trans}), (\ref{kyoto-omega-1}), (\ref{kyoto-omega-2}).  Moreover, in the case of 
trivial background $\hat{F}(A)=0$ such a singular gauge transformations  are allowed and produce no
action.  A crucial question is whether the strings (\ref{Sigma-colored}) are equivalent to gauge
transformations when quantum fluctuations are included. Of course, if it were not so that the
gauge transformations considered are singular, there would be no doubt that the quantum corrections
do not destroy equivalence of the two field configurations related by a gauge transformation. But
because of the presence of singularities we performed an explicit analysis of the quantum
corrections. The result is that the quantum corrections do not distinguish between the standard
perturbative vacuum and the zero-action field configuration presented in the preceding section.

For the sake of definiteness we consider a straight Dirac string in the partition function
(\ref{Sigma-path-integral})
\beq
\label{quantum-PF}
Z[\Sigma] = \int \cD A \exp\Bigl\{ - {1\over 4 g^2} \int d^4 x \; \Bigl[ F^a_{\mu\nu} 
+ 4\pi\; \dual{\Sigma}^a_{\mu\nu} \Bigr]^2 \Bigr\}\,,
\eeq
\beq
\label{quantum-Sigma}
\Sigma^a_{\mu\nu} = - 4\pi \; \delta_{0,[\mu} \; \delta_{\nu],3} \;
\delta^{a,3} \cdot \Theta(z) \delta(x)\delta(y)\,.
\eeq
The "classical" solution of the field equations is the pure gauge configuration
(\ref{kyoto-monopole}) $A^{cl}= i \Omega^+ \diff \Omega$ where
$\Omega$ is given by (\ref{kyoto-omega-1}) and the "classical" action is $S^{cl}=0$.
Expanding the action up to the second order in small perturbations $A=A^{cl}+a$ one finds that
in the background gauge $D_\mu(A^{cl})a_\mu=0$ Eq.~(\ref{quantum-PF}) becomes:
\beq
\label{quantum-PF-1}
Z[\Sigma] =  \mathrm{Det}^{-1}[D^2(A^{cl})]
\eeq
since in the present case the Pauli paramagnetic term is zero. With conventional normalization
to the perturbative vacuum to vacuum amplitude the question whether the string
$\Sigma^a_{\mu\nu}$ is relevant on quantum level,
is equivalent to exploring the spectrum of the operator $D^2(A^{cl})$:
\beq
\label{quantum-D}
D^2(A^{cl}) ~=~  M_1 ~+~ M_2 ~+~ M_3
\eeq
\beq
\label{quantum-D-decomposed}
M_1^{ab} = \delta^{ab} \vec{\diff}^2
\qquad
M_2^{ab} = 2 \varepsilon^{akb} \vec{A}^k \vec{\diff}
\qquad
M_3^{ab} = \varepsilon^{akb} \vec{\diff} \vec{A}^k + \vec{A}^a \vec{A}^b
	- \delta^{ab} \vec{A}^k \vec{A}^k
\eeq
where superscripts denote the color indices and vector notations are used for spatial components of $A^{cl}$.
Using the explicit form (\ref{kyoto-omega-1}) one finds that the non-zero elements of the antisymmetric matrix
$M_2$ are
\bea{rcl}
\label{quantum-M_2}
M_2^{12} & ~=~ &
	\frac{2}{r^2} \; {1+\cos\theta \over \sin^2\theta} \;\diff_\varphi \\
\rule{0mm}{6mm}
M_2^{13} & ~=~ & 
	-\frac{2}{r^2}\;\left(\cos\varphi\;\diff_\theta-{\sin\varphi\over\sin\theta}\;\diff_\varphi\right) \\
\rule{0mm}{6mm}
M_2^{23} & ~=~ &
	-\frac{2}{r^2}\;\left(\sin\varphi\;\diff_\theta+{\cos\varphi\over\sin\theta}\;\diff_\varphi\right)
\eea
where $\theta$ and $\varphi$ are the polar and azimuthal angles, respectively. In the same coordinate
system the matrix $M_3$ is given by
\beq
\label{quantum-M_3}
M_3 = - {2\over r^2} \; {1+\cos\theta \over \sin^2\theta} \;
\left[
\begin{array}{ccc}
	1 &
	0 &
	0 \\
	0 &
	1 &
	0 \\
	-\cos\varphi \; \sin\theta  &
	-\sin\varphi \; \sin\theta  &
	1-\cos\theta \\
\end{array}
\right]
\eeq
It is convenient to perform the transformation $D^2(A^{cl}) \to R\; D^2(A^{cl}) \; R^{-1}$
where the matrix $R$ transforms to the spherical basis:
\beq
\label{quantum-R}
R = \left[ \begin{array}{ccc}
	\cos\varphi \sin\theta &
	\sin\varphi \sin\theta &
	\cos\theta \\
	\cos\varphi \cos\theta &
	\sin\varphi \cos\theta &
	-\sin\theta \\
	-\sin\varphi &
	\cos\varphi &
	0 \\
\end{array} \right]\,.
\eeq
One finds that in the new basis:
\beq
\label{quantum-D-2}
D^2(A^{cl}) = \vec{\diff}^{\;2} + {1\over r^2} \cdot
\left[ \begin{array}{ccc}
	0 &
	0 &
	0 \\
\rule{0mm}{6mm}
	0 &
	-{1 \over \sin^2\theta} &
	{2 \over \sin^2\theta} \diff_\varphi \\
\rule{0mm}{6mm}
	0 &
	-{2 \over \sin^2\theta} \diff_\varphi &
	-{1 \over \sin^2\theta} \\
\end{array} \right]\,.
\eeq
Next, introduce
\beq
\label{quantum-new-basis}
a_0 ~=~ a_r \qquad\qquad a_{\pm}  ~=~  - {i \over \sqrt{2}}\;(a_\theta \mp i a_\varphi)\,,
\eeq
then the operator  $D^2(A^{cl})$ becomes diagonal:
\beq
\label{quantum-D-final}
D^2(A^{cl}) ~=~ \vec{\diff}^{\;2} + 
{1\over r^2} \cdot \mathrm{diag}\left[ 
0\;,\; -{1 \over \sin^2\theta}(1-2i \diff_\varphi) \;,\; -{1 \over \sin^2\theta}(1+2i \diff_\varphi)
\right]
\eeq
Once $D^2(A^{cl})$ is brought to the diagonal form, the direct calculation
shows that the spectrum of (\ref{quantum-D-final}) is identical to that
of free Laplacian $\vec{\diff}^{\;2}$. Therefore, the quantum
fluctuations do not distinguish the string,
Eq.~(\ref{quantum-PF},\ref{quantum-Sigma}), from the perturbative
vacuum. Thus the modified theory
(\ref{part-fun-general},\ref{new-action}) is perturbatively equivalent
to the conventional gluodynamics.

\section{Strings in General Background.}
\label{Strings-in-background}
\noindent
We have shown that in the perturbation theory both closed and open Dirac strings with arbitrary color orientation
carry no action and are thus pure gauge artifacts. On one hand, this conclusion is welcome since it shows that the
continuum limit as understood in this paper perturbatively is the same as the standard continuum limit.
And, indeed, there are no doubts in the validity of the standard  perturbation theory. On the other hand, if it
were so that the singular fields admitted now into the continuum formulation are not associated with any action
at all then the new formulation would be equivalent to the standard one. 

In this section we address this issue on a non-perturbative level and imitate the non-perturbative fields by a
smooth background. The crucial observation then is that only the Dirac strings with the proper color alignment
(\ref{Sigma-colored}) cost no action in the continuum limit,
while the Dirac strings associated with the monopoles defined a la 't~Hooft \cite{tHooft-abelian-gauges}
do not satisfy this constraint.

Consider as an example the class of Abelian gauges of  Ref.~\cite{tHooft-abelian-gauges} which are
defined by the requirement that some adjoint operator $h^a$ is to be directed  along $\tau^3$ in the
color space. This operator may be arbitrary in principle, but for the given $h^a$ the remaining gauge
freedom  consists of U(1) rotations around $\tau^3$:
\bea{ccc}
\label{tHooft-gauge-transformation}
 & \Omega^+ \hat{h} \Omega ~\to ~ h^3 \tau^3\,, & \\
& & \\
\Omega ~=~ \tilde{\Omega}\;H\,, \qquad &
\tilde{\Omega} \in G/U(1)\,, &
\qquad H\in U(1)\,.
\eea
This gauge condition is not defined on the set of points where $h^a=0$ which in four dimensions defines
the monopole trajectory. Clearly enough, the Dirac strings associated with the monopoles are oriented
along the third direction in the color space. Thus, there exist now two different directions in the
color space determined by the background field and through the gauge fixation inherent to the
definition of the monopoles.  

Note that while the boundary of the string singularity is fixed by the equations $h^a=0$, the  actual
position of the string may be changed with a suitable choice of $H$.  Indeed, the gauge transformation
(\ref{tHooft-gauge-transformation}) gives rise to the Dirac string:
\beq
\label{tHooft-string}
\hat{F}(A^\Omega) ~=~ H^+ \tilde{\Omega}^+ \hat{F}(A) \tilde{\Omega} H
+ i H^+ \left[\diff , \diff \right] H
+ i H^+ \left(\tilde{\Omega}^+ \left[\diff , 
\diff \right]\tilde{\Omega} \right) H.
\eeq
The both terms $H^+ \left[\diff , \diff \right] H$ and
$\tilde{\Omega}^+ \left[\diff , \diff \right]\tilde{\Omega}$ 
are proportional to $\tau^3$. Let us stress that the  freedom to choose the U(1) matrix $H$ allows to
shift the position of the string in the ordinary space. Simultaneously, the background field is also
transformed and one may not say that shifting the Dirac string brings no change in the action.
However, since there is no spontaneous breaking of the color symmetry, the dependence on the position
of the string drops off after integrating over all the background fields.

Note that similar considerations apply to the  't~Hooft loop operator which we consider in the next
section. Indeed, the definition of the 't~Hooft loop operator as well as its value in a given
background are string dependent. But the freedom to shift the position of the string in the path
integral approach guarantees that no physical result depends on the string position.

It is amusing to note that the present considerations provides with a general framework to understand
the correlation between instantons and monopoles which has been discussed in various contexts recently
(see, e.g., \cite{gubarevchernodub}). Indeed, background fields in the physical vacuum are described
realistically by instantons (see \cite{Schuryak-review} for a review).
On the other hand , monopoles, as is argued above, are meaningful only in the presence of background
fields.

\section{The 't~Hooft Loop in the Continuum Limit.}
\label{tHooftSc}
\noindent
In this Section we show that the construction presented above allows to define and study
the properties of the 't~Hooft loop operator in the path integral formalism.

The 't~Hooft loop in SU(2) LGT with the one-plaquette action (\ref{LGT-general}) has the following form:
\beq
\label{lattice-tHooft-loop}
H_{lat}(\Sigma_j) =
\exp\Bigl\{ \frac{4}{g^2} \sum\limits_{p \in \dual{\Sigma}_j}
\Bigl[S_P\Bigr(1-\Tr{2} U[\diff p] \Bigl) - S_P\Bigl(1+\Tr{2} U[\diff p] \Bigr)\Bigr]\Bigr\}\, ,
\eeq
where $\dual{\Sigma}_j$ is the set of the plaquettes dual to the surface $\Sigma_j$ with the boundary
$j$. In the path integral formulation the 't~Hooft loop effectively changes the sign of the plaquette
variables $U[\diff p]$ belonging to $\dual{\Sigma}_j$: $U[\diff p] \to - U[\diff p]$. To define the
't~Hooft loop in the continuum we consider the path integral, Eq.~(\ref{Sigma-path-integral}), with an open
orientable non self-intersecting surface $\Sigma^a_j = n^a \Sigma_j$, $\diff \Sigma_j = j$, multiplied
by the Wilson loop $W_J(\cC)$:
\beqn
\label{Wilson-loop-part-fun}
Z(\cC, \Sigma_j) & = & \int \cD A \exp\left\{ - {1\over 4 g^2} \int d^4 x \;
\left[ F^a_{\mu\nu} + 4\pi q\; \dual{\Sigma}^a_{j\;\mu\nu} \right]^2 \right\} \cdot W_J(\cC)\,,\\
\label{Wilson-loop}
W_J(\cC) & = & \tr \Pexp \Bigl\{i\oint\limits_\cC T^a_J  A^a_\mu dx_\mu \Bigr\}\, ,
\eeqn
where $T^a_J$ are the generators of SU(2) in the representation $J$. It is convenient to use the
following integral representation~\cite{kirillov}:
\beqn
W_J (\cC) =  \int \! \cD \omega \exp\Bigl\{ i J \oint\limits_\cC
	\left[ A^\omega_\mu \right]^3 dx_\mu\Bigr\}\,,
\quad
\Bigl[ A^\omega_\mu \Bigr]^3 = \tr\left[\tau^3 \; \omega^+( A_\mu+i \diff_\mu) \omega
\right]\,, \nonumber
\eeqn
where the path integral is over all gauge transformations $\omega$ of the potential $A$ on the
contour $\cC$. Therefore
\beq \label{Wilson-loop-part-fun-1}
Z(\cC , \Sigma_j)= \int \cD A \cD \omega \; \exp\{ - S(A,\Sigma_j)
+ i J \oint\limits_\cC  \left[ A^\omega_\mu \right]^3 dx_\mu \}\,.
\eeq
The action $S(A,\Sigma_j)$ and the measure of integration $\cD A$ are gauge invariant.
The gauge transformation $A ~\to ~ A^{\omega^{-1}}$ defined on $\cC$ allows to factorize the
integral $\cD \omega$:
\beq
\label{Wilson-loop-part-fun-2}
Z(\cC , \Sigma_j)= \int \cD \omega \;\cdot\;\int\cD A\exp\{-S(A,\Sigma_j)
+ i J \oint\limits_\cC A^3_\mu dx_\mu\}\,.
\eeq
The expression (\ref{Wilson-loop-part-fun-2}) has the following meaning: if there is no gauge fixing
in the path integral (\ref{Wilson-loop-part-fun}) the Wilson loop may be calculated exactly by
restricting the gauge potential $A$ to diagonal U(1) subgroup of SU(2)~\cite{kirillov}.

Now we deform the surface $\Sigma_j$ spanned on the contour $j$ to another orientable non
self-intersecting surface $\Sigma'_j$ spanned on the same contour. We also consider the field
$n'^a(\sigma)$ defined on $\Sigma_j'$ according to (\ref{Sigma-colored}). Then the closed surface
$\cS = \Sigma_j - \Sigma'_j$, which bounds the 3-volume $\cV_\cS = \cV_{\Sigma_j - \Sigma'_j}$, has no
self-intersection points and there exists a vector field $n^a(x)$, defined in the whole space--time,
\bea{ccc}
\label{n-a-on-surface-1}
n^a(x) = n^a(\tilde{x}) & \mbox{for} & x\in \Sigma_j\,, \\
n^a(x) = n'^a(\tilde{x}) & \mbox{for} & x\in \Sigma'_j\,.
\eea
In particular, $n^a(x)$ is defined on the contour $\cC$ and there exists an SU(2)
matrix $h \in \cC$, such that:
\beq
\label{3-to-n-a}
\left[ h \; \sigma^3 \; h^+ \right]^a = n^a(x) \qquad \qquad x\in \cC\,.
\eeq
After the gauge transformation $A~\to~A^h$ Eq.~(\ref{Wilson-loop-part-fun-2}) becomes:
\beq
\label{Wilson-loop-part-fun-3}
Z(\cC , \Sigma_j)= \int \cD \omega \;\cdot\;\int\cD A \exp\{ - S(A,\Sigma_j) 
+ i J \oint\limits_\cC (  n^a A^a + [ i h^+ \diff h]^3 )
\eeq
Consider now the additional gauge transformation $\Omega(\cV_\cS)$
(\ref{closed-gauge-trans},\ref{closed-gauge-trans-param}) with
$\cV_\cS = \cV_{\Sigma_j - \Sigma'_j}$. A straightforward calculation
gives
\beq
\label{gauge-1}
n^a \left[A^\Omega_\mu\right]^a = n^a A^a_\mu - 4\pi q\; V_\mu \,.
\eeq
As is shown in Section~\ref{String-independence} for integer and half-integer charges $q$ this
gauge transformation shifts $\Sigma_j$ to $\Sigma'_j$:
\beq
\label{gauge-2}
S(A^\Omega, \Sigma_j) = S(A,\Sigma'_j)\,.
\eeq
For self--consistency of the theory, the Wilson loop is to be invariant under the gauge
transformations. If we apply the transformation (\ref{gauge-1},\ref{gauge-2}) to $Z(\cC , \Sigma_j)$,
see Eq.~(\ref{Wilson-loop-part-fun}), we get:
\beq
\label{commutator-Z}
\Omega(\cV_\cS):\quad
Z(\cC , \Sigma_j) \to Z(\cC , \Sigma_j - \cS) \cdot
e^{-i \; 4\pi q J \; \cL(\cC,\cS)}\,,
\eeq
where $\cL(\cC,\cS)$ is the 4D linking number between the closed contour $\cC$ and the closed
surface $\cS$:
\beqn
\label{linking-number}
\cL(\cC, \cS) =
\oint\limits_{\cS} \left(\dual{d^2\sigma}\right)_{\mu\nu}
\oint\limits_{\cC} dx_\nu
\diff_\mu \Delta^{-1}(\tilde{x}(\sigma) - x) \, .
\eeqn
Since $\cL \in \Z$ and $J$ takes integer and half--integer values the independence of the Wilson loop
on the gauge transformations $\Omega$ implies the quantization condition:
\beqn
\label{quantum-condition}
q \in \Z\,.
\eeqn
This equation is a direct analog of the Dirac quantization condition in electrodynamics. Physically
it means that the electrically charged particle introduced by Wilson loop does not scatter on the Dirac
string $\cS$.

Now we show that the 't~Hooft loop operator $H(\Sigma_j)$ is given by:
\beqn
\label{t'Hooft-loop-operator}
H(\Sigma_j) = \exp\Bigl\{ S(\hat F) - S(\hat F + 2 \pi
\dual{\hat \Sigma}_j) \Bigr\}\, ,
\eeqn
where the surface  $\Sigma^a_j$ is bounded by the contour $j$ and is
given by Eq.~(\ref{Sigma-colored}), the action $S$ is defined by
Eq.~(\ref{new-action}). Indeed, the transformation (\ref{commutator-Z}), when applied to
the quantum average of the product of the fundamental, $J= 1
\slash 2$,  Wilson loop and operator \eq{t'Hooft-loop-operator}, gives:
\beq 
\label{canonical-commutator}
<H(j , \Sigma_j) \; W_{1/2}(\cC)> =< H(j ,\Sigma'_j ) \; W_{1/2}(\cC)
\cdot e^{ i \pi \cL (\cC, \Sigma_j -\Sigma'_j) }> \,.
\eeq
This formula proves that the operator $H$ is the 't~Hooft loop
operator since it is in accordance with relations given in
Refs.~\cite{tHooft-tHooft-loop,Polchinski-82}.

\section{Predictions for the 't~Hooft loop.} 
\label{Monopole-interaction} 
\noindent 
In this Section we consider the rectangular $T \times R$ time-like
contours $j$, Eq.~(\ref{t'Hooft-loop-operator}), with $T \gg R$. Then
the expectation value of the 't~Hooft loop operator is 
\beq
\label{potential-general} <H(\Sigma_j)> = <H(T,R)> ~\sim ~ e^{-T V_{m\bar{m}}(R)}\,, 
\eeq
where by analogy with the Wilson loop we refer to the quantity
$V_{m\bar{m}}(R)$ as to the intermonopole (monopole--antimonopole)
potential. It is worth emphasizing that the potential $V_{m\bar{m}}$
corresponds to the $|Q_M|=1$ monopoles while in Ref.~\cite{Suganuma-98} the monopole--antimonopole
potential with the charge $|Q_M|=2$ has been studied.
These double charged monopoles are identified with the Abelian monopoles in Abelian projections.

Below we formulate predictions for the 't~Hooft loop operator,
Eq.~(\ref{t'Hooft-loop-operator}), and its expectation value,
Eq.~(\ref{potential-general}). In particular, we show that the
't~Hooft loop operator inserts the pair of $|Q_M|=1$ monopoles which
are pure Abelian in the Maximal Abelian gauge.  This fact allows to
fix the short distance asymptotic of the intermonopole potential.  We
argue then that this potential at larger distances at zero and high
temperatures is of Yukawa type.  We also find the screening mass in
both cases and compare it with the masses measured on the lattice
\cite{Rubakov}. Our estimates turn to be in agreement with the
numerical data.

\subsection{Intermonopole Potential at Small Distances.}
\label{tHooft-loop-small-distances}
\noindent
Consider the potential $V_{m\bar{m}}(R)$ at small distances for the
mo\-no\-po\-le--anti\-mo\-no\-po\-le pair introduced by the operator $H(T,R)$. The definition
(\ref{t'Hooft-loop-operator}) shows that we have enough gauge freedom to take
$\Sigma^a_j = \delta^{a,3} \Sigma_j$ on the non self-intersecting surface $\Sigma_j$.
Then at the classical level the solution of the corresponding equations of motion is \cite{Samuel-83}:
\bea{c}
\label{small-R-solution}
A^3_\mu dx_\mu = {1\over 2}\left( {z_+ \over r_+} - {z_- \over r_-}\right) \; d\varphi\,,
\qquad
A^{1,2}_\mu = 0\,,
\\
\rule{0mm}{1.5\bigskipamount}
z_\pm  = z \pm R/2\,,
\qquad
\rho^2 = x^2 + y^2\,,
\qquad
r^2_\pm = z^2_\pm + \rho^2\,,
\eea
and represents the Abelian monopole-antimonopole pair separated by the distance $R$. Since
the monopoles in (\ref{small-R-solution}) have minimal allowed magnetic charge $q=1/2$
(see Section~\ref{String-independence}), at the classical level the intermonopole potential
is given by:
\beq
\label{small-R-potential}
V_{m\bar{m}}(R) ~=~ -{\pi \over g^2\;R}
~=~ - \pi^2\;\beta \;{1\over 4\pi R}\,, \qquad \beta = {4 \over g^2} \, .
\eeq
Note that the statement on the Coulombic nature of the intermonopole potential at short
distances is well known \cite{Samuel-83,Rubakov}. However, the fixation of the coefficient in front of
$1/R$ is new, to the best of our knowledge\footnote{
The same coefficient is derived in Ref.~\cite{Rubakov-last}, which appeared on the day
of submission of the present paper.
}.

Since the potential (\ref{small-R-potential}) was obtained for pure Abelian fields, 
we still have to prove that the general solution with minimal energy in SU(2) gluodynamics
is indeed a gauge rotation of (\ref{small-R-solution}).
A straightforward way to test the Eq.~(\ref{small-R-potential})
is to investigate the problem numerically.
We have calculated the expectation value of the 
't~Hooft loop in the standard SU(2) lattice gauge theory in the limit $\beta \to \infty$.
Technically this limit is realized with the help of the so--called cooling procedure which was used to 
minimize the expectation value of the 't~Hooft loop with respect to the classical lattice
equations of motion.
Our calculations have been performed on the three-dimensional $24^3$ lattice with periodic
boundary conditions, which is adequate to consider the static mo\-no\-po\-le--anti\-mo\-no\-po\-le pair.
We minimized the 't~Hooft loop operator, which
creates static mo\-no\-po\-le\ and anti\-mo\-no\-po\-le separated by the distance $R$.
We have fitted our data for the monopole--antimonopole potential by:
\beq
\label{small-R-potential-lattice}
V^{lat.}_{m\bar{m}}(R) ~=~ - \pi^2 \, \beta \, \Delta^{-1}_{lat.}(R)\,,
\eeq
where $\Delta^{-1}_{lat.}(R)$ is the three-dimensional lattice Coulomb potential.
Eq.~(\ref{small-R-potential-lattice}) is the lattice regularization of the continuum 
expression (\ref{small-R-potential}). Note that the lattice and continuum potentials 
drastically differ from each other and this is of crucial importance in fitting the 
lattice data: the potential (\ref{small-R-potential-lattice}) is regular at $R=0$ contrary
to  (\ref{small-R-potential}).

Our numerical calculations confirmed the behavior (\ref{small-R-potential-lattice})
with accuracy $2\%$. We also
observed that after the cooling procedure the fields are Abelian up to
a gauge transformation. In more detail, we found that in the Maximal
Abelian gauge the gauge fields are diagonal and consist of the Abelian
monopoles located at the boundary of the string $\Sigma_j$,
Eq.~(\ref{t'Hooft-loop-operator}).  Therefore the classical limit of
the state created by non-Abelian 't~Hooft loop is the {\sl Abelian}
monopole--antimonopole pair.

Moreover, once the result (\ref{small-R-potential}) is established classically, the effect of the
quantum corrections is also known on general grounds. Namely, the effect of the quantum corrections
should be reduced to the replacement of the bare coupling by the running one, $g^2\to g^2(R)$.
Although the result is easy to guess, its derivation might look rather mysterious. Indeed, we have
now both non-Abelian magnetic monopoles as external objects and ordinary gluons as virtual
particles. At first sight we need both the standard and dual formulations of the gluodynamics to
describe interaction both with magnetic and electric charges. While in case of U(1) gauge theories
such a formulation is well known \cite{Zwanziger}, it is absent in case of non-Abelian theories.
Thus, we seem to know how the coupling runs although do not know, whose coupling is it! 

We think that the resolution of the paradox is in the Abelian nature of the $|Q_M|=1$ monopoles
established above. Indeed, the classical considerations allow us to fix vertices, or the
Lagrangian. The exact Abelian nature of the monopoles implies that once we choose an Abelian gauge
fixing only neutral bosons (diagonal gluons)
interact with the monopoles $|Q_M|=1$. The charged vector bosons are
still manifested through the loops.  Thus, the situation is similar to the U(1) case with inclusion
of the effect of virtual charged particles.  As for the virtual monopoles, their effect can be
neglected since the monopoles $|Q_M|=1$ are infinitely heavy in the continuum limit. There is no
much difficulty to deal with this problem and one can check that indeed the effect of the loops is
the running of the coupling $g^2$. The details of the U(1) case can be found in the review in Ref.
\cite{Zwanziger}, see also the recent paper \cite{Nielsen-99}.  As for the perturbative
calculations in non-Abelian theories in the Abelian projections, they can be found in Ref.
\cite{physrev}

\subsection{Abelian Dominance and Intermonopole Potential.}
\label{Abelian-Dominance}
\noindent
Next we discuss the  monopole--antimonopole potential at larger distances. The basic idea is to apply
the Abelian Dominance hypothesis \cite{Abelian-Dominance}. Indeed, as has been shown above
the 't~Hooft loop operator inserts the $|Q_M|=1$ monopole pair in the vacuum of SU(2) gauge theory.
Moreover, in the Maximal Abelian gauge these monopoles become a pure Abelian
objects. Therefore it is natural to expect 
that in this particular gauge the dominant contribution to the potential
(\ref{potential-general}) is due to the interaction with Abelian fields.
In the Maximal Abelian gauge the vacuum of zero temperature SU(2) gluodynamics is a dual superconductor
where, instead of condensate of Cooper pairs, there exists a monopole condensate.
The principle of Abelian Dominance assumes that long distance properties of gluodynamics
might be explained in terms of the interaction with the monopole condensate
(for reviews see, e.g., \cite{reviews}).

Following this logic, we expect that at the zero temperatures the monopole--antimonopole potential is:
\beq
\label{large-R-potential}
V_{m\bar{m}}(R) ~=~ - {\pi\over g^2} \;{ e^{-\mu R}\over R}
\eeq
\beq
\label{large-R-potential-lattice}
 V^{lat.}_{m\bar{m}}(R) ~=~ - \beta \; \pi^2  \;
(-\Delta + \mu^2)^{-1}_{lat.}(R)
\eeq
where $\mu$ is the dual photon mass $m_V$ and $(-\Delta + \mu^2)^{-1}_{lat.}$ is the three-dimensional
lattice Yukawa potential.  The recent numerical investigation of the 't~Hooft loop in SU(2) lattice
gauge theory \cite{Rubakov} agrees with Eq.~(\ref{large-R-potential}).
The value of $\mu \approx 3.24(42)\sqrt{\sigma}$ obtained in Ref.~\cite{Rubakov} is quite 
close to the dual photon mass $m_V \approx 1\,\,{\mathrm{GeV}} = 2.3 \, \sqrt{\sigma}$
found in Ref.~\cite{MIP}.
Let us also note that we would not identify directly the mass $\mu$ in Eq.~(\ref{large-R-potential})
with a glueball mass.
Indeed, the definition of the 't~Hooft loop is highly nonlocal and includes
a Dirac string with infinite action. Therefore, the validity of the dispersive relations
is questionable in this case. Note, however, that $\mu$ in Eq.~(\ref{large-R-potential})
coincides with the $0^{++}$ glueball mass in the strong coupling expansion \cite{Samuel-83}.
If this result is valid also in the weak coupling limit, then the Abelian Dominance is reduced
to the prediction that the dual photon mass $m_V$ coincides with the $0^{++}$ glueball mass.
Comparison of numerical results for the Yukawa mass $\mu$ with glueball masses can be found
in \cite{Rubakov}. 

Note that the prediction (\ref{large-R-potential}) is highly
non-trivial in fact. Indeed the $|Q_M|=1$ monopoles are so to say
fundamental monopoles which look as Abelian monopoles at short
distances and are associated for this reason with an infinite
action. They are introduced therefore as external objects via the
't~Hooft loop, similar to introduction of infinitely heavy quarks via
the Wilson loop. The $|Q_M|=2$ monopoles, on the other hand, have a
finite action and their description as a fundamental objects seems to
be granted only at large distances. This could be manifested, in
particular, through existence of an intermediate region between the
distances where the Coulombic and Yukawa pictures apply. In other
words, the coefficient in front of the Coulombic term could have not
matched the coefficient in front of the Yukawa-like
potential. However, existing data about the 't~Hooft loop
\cite{Rubakov} indicate that the matching is exact, within the error bars.
In other words, the dual Abelian Higgs model of QCD vacuum works
already at smallest distances available on the lattice. Similar
conclusions can be drawn in fact from the studies of the heavy quark
potential induced by monopoles \cite{Suzuki} and from description of
the structure of the flux string \cite{MIP,Bali-string-structure}, for
a review see
\cite{GuPoZa-99}.

\subsection{Finite Temperatures.}
\noindent
The authors of Ref.~\cite{Rubakov} have also performed numerical calculations of the 't~Hooft
loop at finite temperatures, and determined the dependence of the Yukawa mass $\mu$ on the temperature.
To provide a theoretical framework for the behavior of the 't~Hooft loop at high temperatures
we can use again the idea of the Abelian Dominance.

In more detail, we estimate the screening mass $\mu$ using the fact that the Abelian model which
corresponds to the high temperature SU(2) gluodynamics is the 3D compact U(1) theory.
Therefore the intermonopole potential at high temperatures is essentially given by
(\ref{large-R-potential},\ref{large-R-potential-lattice}), with $\mu$ now being the Debye mass
\cite{Polyakov-77}:
\beq
\label{m_D}
m^2_D ~=~ 16\pi\; {\rho \over e^2_3} \,,
\eeq
where  $\rho$ is the density of Abelian monopoles and $e_3$ is the corresponding
three-dimensional coupling constant.
To estimate the temperature dependence of $m_D$ we use
the numerical results of Ref.~\cite{Bornyakov}, where the density of Abelian monopoles was
obtained\footnote{
Note that the original result of Ref.~\cite{Bornyakov} for the lattice monopole density 
is: $\rho_{lat.} = 0.50(1) \, \beta^3_G$, where $\beta^3_G$ is a three dimensional coupling 
constant which is expressed in terms of the 3D electric charge $e_3$ and lattice spacing $a$
as  $\beta^3_G = 4 \slash (a \, e^2_3)$. The physical density $\rho$ of monopoles is given by
$\rho=\rho_{lat.} \, a^{-3}$ which can easily be transformed into Eq.~(\ref{monopole-density}).
}:
\beq
\label{monopole-density}
\rho = 2^{-7} (1 \pm 0.02) \, e^6_3\,,
\eeq
Therefore
\beq
\label{m_D-phys}
m_D = 1.11(2) \, e^2_3 \,.
\eeq
Moreover, at high temperatures we can use the dimensional reduction formalism and express the 
3D coupling constant
$e_3$ in terms of the 4D Yang--Mills coupling $g$. At the tree level one has
\beq
\label{e3}
e^2_3 (T) = g^2(\Lambda,T) \, T\,,
\eeq
where $g(\Lambda,T)$ is the running coupling calculated at the scale $T$,
\beq
\label{e4}
g^{-2}(\Lambda,T) = {11\over 12\pi^2} \log \Bigl({T \over \Lambda} \Bigr)
+ {17\over44 \pi^2} \log \Bigr[2 \log
\Bigl({T \over \Lambda}\Bigr)\Bigr]\,,
\eeq
and $\Lambda$ is a dimensional constant which can be determined from lattice simulations.

At present the lattice measurements of the $\Lambda$ parameter are not very precise. We 
use the results of two particular calculations.
Namely, in Ref.~\cite{Heller} the lattice data for the gluon propagator have been used
to determine the so--called "magnetic mass" in high temperature SU(2) gluodynamics.
These measurements imply the following value of $\Lambda$:
\beq
\label{lh}
\Lambda = 0.262(18)\, T_c = 0.197(14)\, \sqrt{\sigma}\,,
\eeq
where $T_c$ is the temperature of the deconfinement phase transition,
$T_c \approx 0.75 \sqrt{\sigma}$. In Ref.~\cite{Bali}, on the other hand,
the spatial string tension has been calculated and the corresponding value of $\Lambda$
turned to be three times smaller:
\beq
\label{lb}
\Lambda = 0.076(13) \, T_c = 0.057(10) \, \sqrt{\sigma}\,.
\eeq
Collecting Eqs.~(\ref{m_D-phys})-(\ref{lb}) we get predictions for the Debye mass which are
shown in Table~\ref{Table:compare} along  with the values of mass $\mu$ obtained numerically
in Ref.~\cite{Rubakov}. One can clearly see that the predictions and the numerical results are
in agreement within the theoretical uncertainties. There are at least three sources of these
uncertainties. First, the value
of $\Lambda$ is not determined precisely as we already noted. Second, we have used the
dimensional reduction which is supposed to work well only at asymptotically high
temperatures, while only one value  $T = 3.676 \, \sqrt{\sigma} \approx 5\, T_c$ in the
Table~\ref{Table:compare} may be considered as high enough.
Third, as we already noted the lattice and continuum Yukawa interactions are substantially different.
For example, we may treat the 't~Hooft loop quantum average studied in Ref.~\cite{Rubakov}
as a two--point correlator in three spatial dimensions. Then we may use 
the results of Ref.~\cite{Mitrjushkin-95} and relate the value of $\mu$
obtained with the use of the continuum propagator to the correct value,
$\mu^{\mathrm{correct}} \approx {2 \over a} {\mathrm{ArcSinh}}\Bigl({\mu a\over 2}\Bigr)$.
If we apply this correction to the values of $\mu$ in Table~\ref{Table:compare} then
for $a \mu = 2.29(55) $ and $T = 3.676 \, \sqrt{\sigma}$ the correction is essential.
Indeed, we obtain: $\mu^{\mathrm{correct}}\approx14.8(3.5)$, which is quite close to our
prediction with $\Lambda=0.197 \sqrt{\sigma}$.

\begin{table}
\begin{center}
\begin{tabular}{|| c | c || c | c ||}
\hline
\hline
\rule[-0.7\bigskipamount]{0mm}{1.8\bigskipamount}
$T \slash \sqrt{\sigma}$ & $\mu \slash \sqrt{\sigma}$ &
\multicolumn{2}{c||}{$m_D\slash\sqrt{\sigma}$}\\
\rule[-0.7\bigskipamount]{0mm}{1.8\bigskipamount}
 & & $\Lambda=0.197\sqrt{\sigma}$ & $\Lambda=0.057\sqrt{\sigma}$ \\
\hline
\rule[-0.7\bigskipamount]{0mm}{1.8\bigskipamount}
0.460  &  3.24(42) & 5.12 & 2.04 \\
\hline
\rule[-0.7\bigskipamount]{0mm}{1.8\bigskipamount}
1.225  &  4.13(41) & 6.16 & 3.82 \\
\hline
\rule[-0.7\bigskipamount]{0mm}{1.8\bigskipamount}
1.838 & 5.43(59) & 7.67 & 5.12 \\
\hline
\rule[-0.7\bigskipamount]{0mm}{1.8\bigskipamount}
3.676 & 17.3(4.1) & 11.97 & 8.69 \\
\hline
\hline
\end{tabular}
\end{center}
\caption{
The screening mass $\mu$ (see (\ref{large-R-potential},\ref{large-R-potential-lattice})) at
different temperatures, Ref.~\cite{Rubakov}, and our predictions for $m_D$ obtained with different
$\Lambda$, Eqs.~(\ref{lh},\ref{lb}).}
\label{Table:compare}
\end{table}

\section*{Conclusions}
\noindent

We have tried to formulate a theoretical framework which would allow for 
mo\-no\-po\-les in the continuum
version of non-Abelian gauge theories. Indeed, monopoles nowadays are very common field
configuration on the lattice. In the continuum, on the other hand, monopoles appear to be
associated with singular fields and divergent action. 

The key element to introduce monopoles in the continuum is to allow
for Dirac strings. While naively the action associated with the Dirac
strings is infinite, they cost no action at all in the compact U(1)
gauge model~\cite{Polyakov-compact-U1}. Thus, within the continuum
formulation, one has to postulate that there are certain singular
fields which cost no action as well. An alternative representation for
the singular fields are Dirac sheets (see, e.g., Eq.~(\ref{Sigma})
above). In the non-Abelian case, we argued that the continuum version
should admit certain singular or stringy fields without any change in
the action. One can say that the Dirac strings which cost no action
are aligned in the color space with the background, or regular fields.

Once the Dirac strings are admitted into the continuum version of gluodynamics, the end points of
the strings, or monopoles, cost in the perturbative vacuum no action either. This is true both
classically and with account of quantum corrections. And this is in distinction from the U(1) case
where the end points are monopoles with an ultraviolet divergent action. As a result, although the
modified continuum version appears very different from the standard one since it allows for
singular potentials inversely proportional to the coupling, perturbatively the two theories are in
fact equivalent.  Thus, at this point the problem seems to be the other way around. Namely, there
is no difficulty any longer to introduce fields which look as monopoles in terms of Abelian fields
but cost no action and appear as gauge artifacts once the full spectrum of the non-Abelian degrees
of freedom is taken into account. 

The difference between the two formulations becomes manifest once the gauge is fixed
a la 't~Hooft \cite{tHooft-abelian-gauges} and background non-perturbative fields are introduced. The
point is that in presence of the background field only those Dirac strings which are parallel to the
background in the color space are non observable. On the other hand, the
definition of the monopoles in terms of
the topology of the gauge fixing introduces Dirac strings which do not satisfy
this condition. As a result, the action associated with the monopoles is not vanishing
any longer. And the monopoles do emerge as possible fluctuations  with finite action
which are present in the continuum theory modified to incorporate Dirac strings.
It is worth emphasizing that upon integration over the background fields the monopole action does not
depend on the position of the Dirac string but only on the monopole trajectory. 

The machinery to prove the independence on the position of the Dirac string is also all what is needed
to introduce a continuum analog of the 't~Hooft loop operator \cite{tHooft-tHooft-loop}. The continuum
formulation of the 't~Hooft loop is one of the central points of this paper.
Furthermore, we were able to derive both rigorous and model-dependent results for
the behavior of the 't~Hooft loop at zero and high temperature SU(2) gluodynamics.

In terms of physical applications, the picture developed explains in generic terms correlation between
instantons and monopoles (for discussion see, e.g., \cite{gubarevchernodub}).
Also, it was demonstrated that while perturbatively the modified theory allowing for the Dirac strings
is equivalent to the standard one, non-perturbatively they are different. 
This might explain a kind of mystery with the non-perturbative $1/Q^2$ corrections from short distances
which seem to exist phenomenologically but evade, so far, theoretical understanding within the
standard framework (for reviews and further references see \cite{reviews1}).

\section*{Acknowledgments}
\noindent
M.N.Ch. and M.I.P. acknowledge the kind
hospitality of the staff of the Max-Planck Institut f\"ur Physik
(M\"unchen), where the work was initiated. The authors are grateful to V.A.~Rubakov 
for useful discussions. Work of M.N.C., F.V.G. and M.I.P.  was partially supported by grants RFBR
99-01230a, RFBR 96-1596740 and INTAS 96-370.


\begin{thebibliography}{99}

\bibitem{reviews}
M.N.~Chernodub, F.V.~Gubarev, M.I.~Polikarpov, A.I.~Veselov, \PTPS{131}{1998}{309}, hep-lat/9802036;\\
G.S.~Bali, preprint {\it HUB-EP-98-57}, hep-ph/9809351;\\
A.~Di~Giacomo, \PTPS{131}{1998}{161}, hep-lat/9803008;\\
T.~Suzuki, \NPPS{30}{1993}{176};\\
M.I.~Polikarpov, \NPPS{53}{1997}{134}, hep-lat/9609020;\\
R.W.~Haymaker, \PRep{315}{1999}{153}.

\bibitem{TP-solution}
A.~M.~Polyakov, \JL{20}{1974}{194}; \\
G.~'t~Hooft, \NP{B79}{1974}{276}.

\bibitem{baker}
M. Baker, J.S. Ball, F. Zachariasen, \PR{D51}{1995}{1968}.

\bibitem{Polyakov-compact-U1}
A.M.~Polyakov, \PL{59B}{1975}{82}.

\bibitem{tHooft-abelian-gauges}
G.~'t~Hooft, \NP{B190}{1980}{455}.

\bibitem{tHooft-tHooft-loop}
G.~t'Hooft, \NP{B138}{1978}{1}.

\bibitem{Brandt-Neri}
R.A.~Brandt, F.~Neri, \NP{B161}{1979}{253}.

\bibitem{Samuel-83}
S.~Samuel, \NP{B214}{1983}{532}.

\bibitem{monopoles-last}
B.L.G.~Bakker, M.N.~Chernodub, M.I.~Polikarpov, \PRL{80}{1998}{30}

\bibitem{Suganuma-98}
H.~Suganuma, H.~Ichie, A.~Tanaka, K.~Amemiya, \PTPS{131}{1998}{559}.

\bibitem{gubarevchernodub}
M.N.~Chernodub, F.V.~Gubarev, \JL{62}{1995}{100};\\
A.~Hart, M.~Teper, \PL{B371}{1996}{261};\\
R.C.~Brower, K.N.~Orginos, C-.I.~Tan, \PR{D55}{1997}{6313};\\
E.M.~Ilgenfritz, S.~Thurner, H.~Markum, M.~Muller-Preussker, \PR{D61}{2000}{054501};\\
T.C.~Kraan and P.~van~Baal, {\it Phys.~Lett.} {\bf B435} (1998) 389;
{\it Nucl.~Phys.} {\bf B533} (1998) 627.

\bibitem{Schuryak-review}
T.~Schafer, E.V.~Shuryak, {\it Rev. Mod. Phys. }{\bf 70} (1998) 323.

\bibitem{kirillov}
A.Yu.~Alekseev, S.L.~Shatashvili, \MPL{A3}{1988}{1551};\\
D.I.~Diakonov, V.Yu.~Petrov, \PL{224B}{1989}{131};\\
P.~Orland, \IJMP{A4}{1989}{3615}; \\
F.A.~Lunev, \NP{B494}{1997}{433}.

\bibitem{Polchinski-82}
A.~Ukawa, P.~Windley, A.~Guth, \PR{D21}{1980}{1013};\\
J.~Polchinski, \PR{D25}{1982}{3325}.

\bibitem{Rubakov}
C.~Hoelbling, C.~Rebbi and V.A.~Rubakov, report at
17th International Symposium on Lattice Field Theory (LATTICE~99),
hep-lat/9909023; \\ C.~Hoelbling, C.~Rebbi, V.A.~Rubakov,
\NPPS{73}{1999}{527},\\ hep-lat/9809113.

\bibitem{Nielsen-99}
L.V.~Laperashvili, H.B.~Nielsen, \MPL{A14}{1999}{2797}.

\bibitem{Zwanziger}
D.~Zwanziger, \PR{D3}{1971}{343};\\
M.~Blagoevi\'{c}, P.~Sen'janovi\'{c}, \PRep{157}{1988}{233}.

\bibitem{physrev}
H.~Min, T.~Lee, P.Y.~Pac, \PR{D32}{1985}{440}.

\bibitem{Abelian-Dominance}
T.~Suzuki, I.~Yotsuyanagi, \PR{D42}{1990}{4257}; \\
S.~Hioki {\it et. al.}, \PL{B272}{1991}{326}, {\it Erratum-ibid.} {B281} (1992) 416;\\
J.D.~Stack, S.D.~Neiman, R.J.~Wensley, \PR{D50}{1994}{3399}.

\bibitem{MIP}
F.V.~Gubarev, E.M.~Ilgenfritz, M.I.~Polikarpov, T.~Suzuki, \PL{B468}{1999}{134}.

\bibitem{Suzuki}
H.~Shiba, T.~Suzuki, \PL{B333}{1994}{461};\\
G.S.~Bali, V.~Bornyakov, M.~Muller-Preussker, K.~Schilling, \PR{D54}{1996}{2863}.

\bibitem{Bali-string-structure}
K.~Schilling, G.~S.~Bali, and C.~Schlichter, \PR{D51}{1995}{5165};~ \NPPS{42}{1995}{273};~
	\PTPS{131}{1998}{645};~ \NPPS{63}{1998}{519};~ \NPPS{73}{1999}{638}.

\bibitem{GuPoZa-99}
F.V.~Gubarev, M.I.~Polikarpov, V.I.~Zakharov, {\it "Physics of the Power Corrections in QCD"},
hep-ph/9908292.

\bibitem{Polyakov-77}
A.M.~Polyakov, \NP{B120}{1977}{429}.

\bibitem{Bornyakov}
V.~Bornyakov, R.~Grigorev, \NPPS{30}{1993}{576}.

\bibitem{Heller}
U.M.~Heller, F.~Karsch, J.~Rank, \PL{B355}{1995}{511}.

\bibitem{Bali}
G.S.~Bali, K.~Schilling, J.~Fingberg, U.M.~Heller, F.~Karsch,\\
\IJMP{C4}{1993}{1179}; \PRL{71}{1993}{3059}.

\bibitem{Mitrjushkin-95}
J.~Engels, V.K.~Mitrjushkin, T.~Neuhaus, \NP{B440}{1995}{555}.

\bibitem{reviews1}
R.~Akhoury, V.I.~Zakharov, \PL{B438}{1998}{165};\\
F.V.~Gubarev, M.I.~Polikarpov, V.I.~Zakharov, \MPL{A14}{1999}{2039};\\
G.S.~Bali, \PL{B460}{1999}{170};\\
K.G.~Chetyrkin, S.~Narison, V.I.~Zakharov, \NP{B550}{1999}{353};\\
M.N.~Chernodub, F.V.~Gubarev , M.I.~Polikarpov, V.I.~Zakharov, hep-ph/0003006;\\
P.~Boucaud {\it et al}, 
{\it "Lattice Calculation of $1/P^2$  Corrections to $\alpha_s$ and of $\Lambda_{QCD}$ in the MOM Scheme"},
hep-ph/0003020.

\bibitem{Rubakov-last}
Ch.~Hoelbling, C.~Rebbi, V.A.~Rubakov, {\it "Free Energy of an SU(2) Monopole-antimonopole Pair"},
hep-lat/0003010.

\end{thebibliography}
\end{document}